\documentclass[12pt]{article}
\usepackage{a4}
\usepackage{amsfonts}
\usepackage{revsymb}
\usepackage{amsmath}
\usepackage{graphicx}

\begin{document}

\begin{titlepage}

\begin{center}
{\Large\bf From SPS to RHIC: Maurice and the CERN heavy-ion programme}
\vfill

{\bf Ulrich Heinz}\\[0.3cm]

%
%
{Physics Department, The Ohio State University, Columbus, Ohio 43210, USA%
\footnote{Permanent address. Email: {\tt heinz@mps.ohio-state.edu}. Work 
supported by the U.S. Department of Energy under contract DE-FG02-01ER41190.}}\\
and\\
{CERN, Physics Department, Theoretical Physics Division, \\
CH-1211 Geneva 23, Switzerland}

\end{center}
\vfill
\begin{abstract}

Maurice Jacob played a key role in bringing together different groups
from the experimental and theoretical nuclear and particle physics 
communities to initiate an ultrarelativistic heavy-ion collision program 
at the CERN SPS, in order to search for the quark-gluon plasma. I review 
the history of this program from its beginnings to the time when the 
Relativistic Heavy Ion Collider (RHIC) at Brookhaven National Laboratory 
(BNL) started operation. I close by providing a glimpse of the important 
discoveries made at RHIC and giving an outlook towards heavy-ion 
collisions at the Large Hadron Collider (LHC). During Maurice's
life and not least through his perpetually strong influence, relativistic
heavy-ion physics has matured and led to discoveries that radiate into 
many other fields of physics. Heavy-ion physicists owe a great deal to 
Maurice Jacob.

\end{abstract}
\vfill
{Invited talk presented at the "Maurice Jacob Memorial Meeting",
CERN,\\ 11 September 2007}
\vfill
\end{titlepage}

\section{Introduction}

I have been asked to talk at this ``Maurice Jacob Memorial Meeting''
about Maurice's role in relativistic heavy-ion physics. I am a little 
younger than the other speakers, and my first serious contact with
Maurice didn't happen until early 1987 when, as Leader of the CERN Theory
Division, he offered me a junior staff position at CERN. This was just
the time when the CERN heavy-ion programme had finished its first run, 
closely followed by a similar program at the AGS at BNL (where I was 
then working), while BNL scientists were already busy trying to obtain 
approval to build a dedicated heavy-ion collision facility, the 
Relativistic Heavy Ion Collider, at 10 times the energy of the CERN SPS. 
Little did I know that Maurice was far ahead of them and already 
thinking about heavy ions in the LHC, at 30 times the energy of RHIC!
Neither did I accept CERN's offer, nor did I stay at BNL -- instead, 
I joined the faculty at the University of Regensburg where I started to 
build my own heavy-ion theory group. But Maurice made sure that I spent 
many springs and summers at CERN, as a regular visitor and a member of 
the SPS program committee, and during these times we, together with 
Leon van Hove, spent endless hours discussing the new heavy-ion data 
and their possible interpretations. It was a heady time of excitement, 
confusion and discovery as we were groping our way through the 
complexities of the theoretical problems posed by the extraordinarily 
tiny, short-lived, extremely dense and highly dynamical ``fireballs'' 
created in these nuclear collisions. In 1998 I went on a 3-year leave 
of absence from Regensburg to join the CERN TH staff and coordinate 
their heavy-ion theoretical activities just when the second phase of the 
SPS heavy-ion programme, the lead beam programme, burst into full 
bloom. As it turned out, I never returned to Regensburg, but instead 
followed a call from the US in late 2000 when I joined the faculty of 
The Ohio State University just after RHIC had finally been completed 
and finished its first run. 

When I came to CERN, Maurice had already officially retired (which 
simply meant he was travelling more in his various international 
leadership roles on which others report elsewhere in this volume). 
But whenever he was home, he looked me up to ``talk physics''. As 
the completion of RHIC drew closer, CERN heavy-ion physicists got 
together to try to assess the results of the CERN heavy-ion programme 
as it was winding down and before the limelight would move across 
the atlantic. As the discussions heated up (see below), it was 
entirely natural that Maurice was ``called back'' from retirement 
and asked to write (together with me) a ``White Paper'' summarizing 
the achievements and insights made at the SPS. A special seminar and 
press conference followed on Feb. 10, 2000, where CERN announced 
\dots\  --- but I am getting ahead of myself! Let us start at the 
beginning \dots

\section{The SPS heavy-ion programme 1986-2003}

Heavy ion collisions at relativistic energies ($E_\mathrm{beam} > A\,m_N$)
were first explored in the mid 1970s at the BEVALAC at Berkeley, motivated
by curiosity about the properties of nuclear matter at densities much
above that of atomic nuclei and by theoretical speculations about 
abnormal states of matter at several times nuclear equilibrium density
\cite{Lee:1974ma}. As QCD, the modern theory of strong interactions, 
became more widely known, it was discovered that it predicted a phase
transition from hadronic matter to a plasma of color-deconfined quarks
and gluons, the quark-gluon plasma, at high densities and temperatures 
\cite{Collins:1974ky}. But the beams provided by the BEVALAC were not
energetic enough to reach this new state of matter. How to get higher
energy heavy ion beams without breaking the bank? Scientists started 
looking at the possibility of injecting heavy ions into existing higher
energy proton accelerators, such as the AGS at Brookhaven and the CERN SPS.
Maurice played a decisive role at CERN in bringing together different
groups, experimentalists and theorists, from the nuclear and particle
physics communities to initiate an ultra-relativistic ($E_\mathrm{beam}\gg
A\,m_N$) heavy-ion collision program at CERN, and to secure the support of 
the laboratory management for this adventure at a time when CERN was 
building LEP. The key move was the proposal to build the heavy-ion program 
using existing equipment not only for the accelerator, but also for the 
ion source and detectors. In 1986 the SPS delivered for the first time 
$^{16}$O beams of 60\,$A$\,GeV to a handful of expe\-ri\-ments, followed in 
subsequent years by $^{32}$S beams at 60 and 200\,$A$\,GeV which were 
used to study S+S, S+U and S+emulsion collisions, as well as a few other 
target nuclei. This phase lasted from 1986 to 1993.

The collisions produced unprecedented large numbers of secondary 
particles whose energy spectra showed tantalizing evidence for thermal
multi-particle production at high temperatures, with an enhanced 
probability for creating strange hadrons. But since the experiments
were using second hand equipment they were not optimized for the task.
It proved difficult to test theoretical predictions with experiments
where different observables were measured in different regions of 
phase space, and the limited experimental acceptance required large 
acceptance corrections which made it tedious to compare data between 
different experiments where there acceptances overlapped. Every exciting
observation (and, as expected when you explore uncharted territory, there
were many!) thus raised more questions than it answered. Furthermore,
sulfur nuclei were kind of small ``heavy'' ions. 

Scientists thus started building a new ion source for lead beams and a
second generation of large, dedicated multi-purpose heavy-ion detectors, 
which went into operation in 1994 with a 160\,$A$\,GeV beam of $^{207}$Pb. 
The CERN Pb-beam program lasted almost 10 years, finishing in 2002 with a 
low energy run at 40, 30 and 20 GeV/nucleon. When the SPS was running
protons instead of heavy-ions, the experiments collected $pp$ and
$pA$ reference data. A final heavy-ion run with an indium beam, which 
the NA60 experiment used to study $^{115}$In+In collisions at 160\,$A$\,GeV, 
yielded an outstanding set of data on electromagnetic signatures from the 
collision fireball that are even now keeping a spotlight on CERN while 
much of the attention of the heavy-ion community has moved to RHIC. 

I looked up on the CERN web site the list of heavy-ion experiments 
completed during this 15-year period. I counted 25 large-detector
experiments (including 7 big multi-year efforts) plus another 20 or so
emulsion experiments. (A parallel effort at the BNL AGS with beam 
energies ranging from 2 to 15 GeV/nucleon, providing $^{28}$Si beams
from 1987 and $^{197}$Au beams from 1995 to 1999, was the second leg
of the relativistic heavy ion program during the final decades of the 
last century.) Clearly, the CERN heavy-ion program (which without 
Maurice might never have happened) ended up being a very strong and 
successful operation which broadened CERN's research portfolio and 
strengthened its reputation as the leading accelerator facility in the 
world. And without the experience gained in the CERN heavy-ion program, 
RHIC could never have become the immediate smashing success that it did.    

The incoming SPS data were new and exciting. We saw more than a 
1000 produced particles per central collision event. Each event was its 
own statistical ensemble, exhibiting collective dynamics -- 
thermodynamics and hydrodynamics became the standard language for 
talking about the collision dynamics. Compared to proton-proton collisions,
strange particle production was enhanced while $J/\psi$ mesons and
other charmonium states were suppressed. The final hadron spectra were
thermal and showed strong, anisotropic collective flow, reflecting 
expansion transverse to the beam direction with more than half the 
speed of light. It was clear that we had succeeded in creating bulk matter 
made of strongly-interacting constituents, and had turned into a new 
breed of condensed matter physicists studying the collective properties 
and phase diagram of such matter!

But, as nice as the data were, their interpretation was difficult: One 
was dealing with a very complex, highly dynamical system of which each
detector measured only a (in some cases small) subset of observables. 
Where measurements overlapped, the data did not immediately agree with
each other. Information how the various interesting observations correlated
with each other was initially quite sketchy. As a result, first opinions 
varied widely, and convergence was slow; in hindsight I can say that
all the right ideas were there, but parsing incorrect concepts and 
interpretations took time.

\section{Maurice, the beacon}

During this time, Maurice's main responsibilities lay elsewhere. As 
recounted by others at this meeting, he was busy directing the CERN
Theory Division, planning for LEP and the LHC, editing journals, books 
and proceedings, and promoting research worldwide in various leading 
roles in the French and European Physical Societies, the European Space 
Agency, etc. Still, as a responsible and supportive father, he
followed the growth of his child, the CERN heavy-ion programme, carefully. 
In continuous discussions with colleagues, and as an active conference 
participant and highly-sought speaker, he tried to distill his own picture
of the heavy-ion fireball evolution from the data. In numerous invited 
opening and summary talks he captured the excitement of the field, 
encouraged the young generation of scientists entering the field in 
droves, and stimulated his colleagues with his insights, speculations 
and proposals. 

Typically, when the field approached a critical transition point, he was
``called upon'' for his assessment and advice. Here is an example from an
opening talk that he gave at the 5th Conference on Nucleus-Nucleus 
Collisions in Taormina in June 1994, when CERN was making the transition
from S to Pb beams and BNL switched from Si to Au beams:
\begin{enumerate}
\item[]
{\small
``There is no doubt that a new state of matter, with a density of at least an
order of magnitude higher than hadronic matter, is created in heavy ion
collisions. We do not yet know what it is and whether or not there is a phase
transition between two very different forms of dense matter, as it is expected
from Quantum Chromodynamics. Nevertheless, since we are after a phase
transition there is nothing like volume and time and also, in order to increase
time, collision energy. Since we have good proven tools now at hand, we can
expect much from an increase in volume and an increase in collision energy.
We can therefore approach with enthusiasm the new rounds of experiments.''
}
\end{enumerate} 
Clearly, Maurice was seeing farther than many of his contemporaries: not
only did he have ``no doubt'' where others were still entrenched in hot
debates, but he was already setting the stage not only for the CERN lead 
beam programme, but also for higher energies at RHIC and LHC!

\section{Preparing the CERN press release}

In 1998, Maurice retired from CERN. The completion of RHIC at BNL was 
looming, with first Au+Au collision at 200 GeV/nucleon {\em in the center 
of mass} (!) planned for 1999. (A series of major leaks in the old 
cryogenic system inherited from the cancelled ISABELLE project delayed
RHIC start-up by another year.) A number of scientists at CERN, 
in particular Reinhard Stock from the University of Frankfurt, pushed
for a public statement by CERN that would summarize the achievements 
and assess the success of the CERN heavy-ion program before RHIC went
into operation. What could (and should) be said at the eve of the RHIC 
era? No doubt the CERN SPS program, and the pioneers who had driven it,
against an adverse tide of scientific and political difficulties, had 
made tremendous advances towards the ultimate goal --- creating 
quark-gluon plasma in the laboratory. This deserved recognition. But 
many detailed questions were still discussed controversially, without 
hope for final resolution at the SPS because its beam energy was 
limited (see below). Ignoring this and claiming unqualified discovery 
of the QGP with the incomplete evidence in hand would do irreparable 
damage to the field (not to speak of the wrath of the colleagues at 
RHIC!). All were acutely aware of the political and sociological aspects
of this assessment.

I joined the CERN Theory Division in August 1998 and was immediately 
drafted to help coordinate the assessment process. In September of 
that year leading participants of the program met for a retreat in
Chamonix. We had about 4-5 months worth of lead beam data to look at 
for our assessment, with more on tape but not yet analyzed. (Typically, 
the heavy-ion runs lasted for 4-6 weeks every year.) The discussions at 
the retreat were contentious and revealed a need for consolidation. We 
went to work. In March 1999 I was asked to give an SPS heavy-ion status 
report to the SPSC; clearly, the lab management had made it a priority
to take stock of the return on their investment.

In late summer 1999, another ``call back to duty'' was issued to Maurice:
he was asked to coordinate, together with me, the writing of a ``White
Paper'' summarizing the scientific achievements of the CERN Pb-beam 
program that reflected the consensus of the community. It was titled
``Evidence for a new state of matter: An assessment of the results from  the
CERN lead beam programme'' \cite{Heinz:2000bk}. I have heard the process
leading to this document described sarcastically as a ``discovery by 
committee''; the truth is that a group of intellectual leaders, in 
perfect scientific tradition, got together and sorted the empirical 
and theoretical evidence to extract a coherent and compelling overall 
picture, throwing out claims and speculations that, based on the 
available facts, didn't hold water. Maurice Jacob was the perfect 
mediator for this process. The resulting White Paper reflected the 
consensus of the SPS heavy-ion community to the extent that consensus 
could be reached; what was not consensus was not written down.

\begin{figure}
\begin{center}
\includegraphics[width=10cm,clip=]{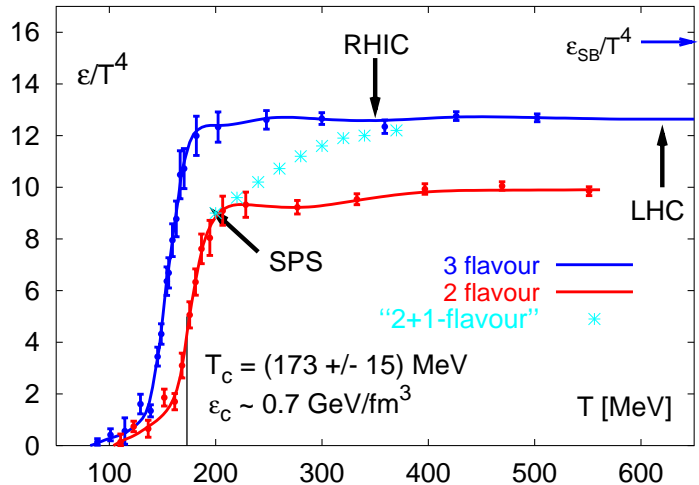}
\end{center}
\caption{\footnotesize
Normalized energy density $\varepsilon/T^4$ vs. temperature $T$ from
Lattice QCD \cite{Karsch:2003jg}, with arrows indicating the temperatures
and energy densities reached at SPS, RHIC and (presumably) LHC. The steep
rise at $T_c$ indicates the phase transition from a hadron gas to the
quark-gluon plasma.
\label{F1}}
\end{figure}

You could ask: Why was this so difficult? The answer is given in 
\cite{Heinz:2000bk}:
\begin{enumerate}
\item[]
{\small
``We emphasize that the evidence collected so far is ``indirect'' since it 
stems from the measurement of particles which have undergone significant 
reinteractions between the early collision stages and their final 
observation. Still, they retain enough memory of the initial quark-gluon 
state to provide evidence for its formation, like the grin of the Cheshire 
Cat in Alice in Wonderland which remains even after the cat has disappeared. 
It is expected that the present ``proof by circumstantial evidence''
for the existence of a quark-gluon plasma in high energy heavy ion 
collisions will be further substantiated by more direct measurements (e.g. 
electromagnetic signals which are emitted directly from the quarks in the 
QGP) which will become possible at the much higher collision energies and 
fireball temperatures provided by RHIC at Brookhaven and later the LHC at 
CERN.''
}
\end{enumerate} 
I should add another complication: As we now know and is shown
in Fig.~\ref{F1}, the SPS beam energy was just barely enough to reach 
the quark-gluon plasma phase: the SPS experiments were ``living
on the edge''. In consequence, the QGP phase lasted only for a very 
short time, converting to hadrons almost immediately, and most of what
was observed were manifestations of collective hadron dynamics. They
camouflaged whatever QGP signatures there were, making it difficult to
extract the latter in an unambiguous fashion.

In the end we wrote ``It walks like a duck, it quacks like a duck, \dots''
but didn't complete the sentence \cite{Heinz:2000bk}:
\begin{enumerate}
\item[]
{\small
``A common assessment of the collected data leads us to conclude that
we now have compelling evidence that a new state of matter has indeed
been created, at energy densities which had never been reached over
appreciable volumes in laboratory experiments before and which exceed
by more than a factor 20 that of normal nuclear matter. The new state
of matter found in heavy ion collisions at the SPS features many of the
characteristics of the theoretically predicted quark-gluon plasma.''
}
\end{enumerate} 
Maurice and I received a lot of sharp-tongued comments about this 
carefully chosen ``politically correct'' formulation. We took it with
humour. What is more important than a never-ending discussion of whether 
the new state of matter for which CERN claimed discovery was the long-sought 
QGP is that, in the process of analyzing the SPS data, we had obtained
a pretty good picture of the time-evolution of the hot fireball
\cite{Heinz:2000bk}:
\begin{enumerate}
\item[]
{\small
``In spite of its many facets the resulting picture is simple: the two 
colliding nuclei deposit energy into the reaction zone which materializes 
in the form of quarks and gluons which strongly interact with each other. 
This early, very dense state (energy density about $3-4$\,GeV/fm$^3$, mean 
particle momenta corresponding to $T\approx 240$\,MeV) suppresses the 
formation of charmonia, enhances strangeness and begins to drive the 
expansion of the fireball. Subsequently, the ``plasma'' cools down and 
becomes more dilute. At an energy density of 1\,GeV/fm$^3$ ($T\approx 
170$\,MeV) the quarks and gluons hadronize and the final hadron abundances 
are fixed. At an energy density of order 50\,MeV/fm$^3$ ($T=100-120$\,MeV) 
the hadrons stop interacting, and the fireball freezes out. At this point 
it expands with more than half the light velocity.''
}
\end{enumerate} 
On 10 February, 2000, CERN held a special seminar \cite{CERNpress}, with
talks from the major SPS heavy-ion experiments, were the discovery of ``A
New State of Matter created at CERN'' was announced. I still have a copy
of a volume put together by the CERN press office that contains press 
clippings from around the world in reaction to this announcement; it is
several hundred pages thick.

\section{From SPS to RHIC}

Even with 20/20 hindsight, there is surprisingly little written in the 
SPS White Paper \cite{Heinz:2000bk} that we would like to take back. Some
specific phrases connected with the discussion of strangeness enhancement
and $J/\psi$ suppression were exaggerated -- data and theory hadn't quite
settled yet when the paper was written. Overall, the first RHIC data
that came in later in the summer of 2000 provided a spectacular
confirmation of the picture that had been developed at the SPS. Of course,
RHIC data went much beyond the observations at the SPS and, as I briefly 
discuss below, led to quite unexpected and exciting discoveries in their 
own right. But it is worth emphasizing that, in doing so, they didn't 
conflict with what had been found at the AGS and SPS, but built naturally 
upon the strong foundations laid at lower energies.

An interesting discussion is connected with Maurice's use of the phrase 
``quarks and gluons roaming freely'' \cite{Heinz:2000bk}:
\begin{enumerate}
\item[]
{\small
``It has been expected that in high energy collisions between heavy 
nuclei sufficiently high energy densities could be reached such that 
this new state of matter would be formed. Quarks and gluons would then 
freely roam within the volume of the fireball created by the collision.
\dots''
}
\end{enumerate} 
It is a powerful phrase that, like the ``grin of the Ceshire Cat'' from
the first quote above, was picked up by journalists around the world in 
their reports on the CERN press release. We meant it to mean ``unshackled, 
freed from their hadronic prison of $\sim 1$\,fm$^3$ living space'', but
scientists (especially ones working at RHIC) frequently (mis)quote it
with the meaning ``moving almost without interactions, as in a dilute gas''.
It cannot be denied that the limitations of QCD perturbation theory for
the description of the transport and flow properties of the QGP were not
yet fully appreciated at the end of the SPS era and, as I will tell, RHIC 
brought about a real paradigm shift. But we definitely knew already before
2000 that the observed strong collective flow of the SPS ``Little Bangs''
required significant interactions among the fireball constituents. 
The White Paper \cite{Heinz:2000bk} therefore makes specific reference to 
``quarks and gluons which strongly interact with each other'' (see above). 
But, when misinterpreted in this particular way, ``roaming freely'' makes 
for a perfect backdrop to highlight the most spectacular discovery made 
at RHIC, namely that of the ``perfect QGP liquid''. 

Before discussing this important RHIC result (that Maurice enjoyed
very much) let me make a few sociological comments about the 
AGS-SPS-RHIC-LHC family. RHIC turned on in June 2000. The RHIC experiments
produced results at a spectacularly fast rate. 20 years of struggle and
consolidation at the SPS and AGS had led to the development of very 
successful and efficient concepts for the RHIC experiments and data 
analysis chains. Many of the key ideas developed at the SPS were 
confirmed at RHIC, enabling the RHIC scientists to quickly move further 
and facilitating the recognition of important new discoveries. Experience 
gained during the SPS program now informs the planning for the LHC 
heavy-ion experiments (ALICE, CMS, ATLAS), and new technologies 
developed for RHIC are flowing back across the Atlantic into the 
construction of LHC detectors. We thus hope for a similarly smooth and 
successful start of the LHC heavy-ion program next year, and I am sure 
Maurice would have loved to see the birth if this great-grand-child of 
his (especially for all the spectacular jets that it will throw about!).
  
\begin{figure}[htb]
\begin{center}
\includegraphics[width=0.463\linewidth,clip=]{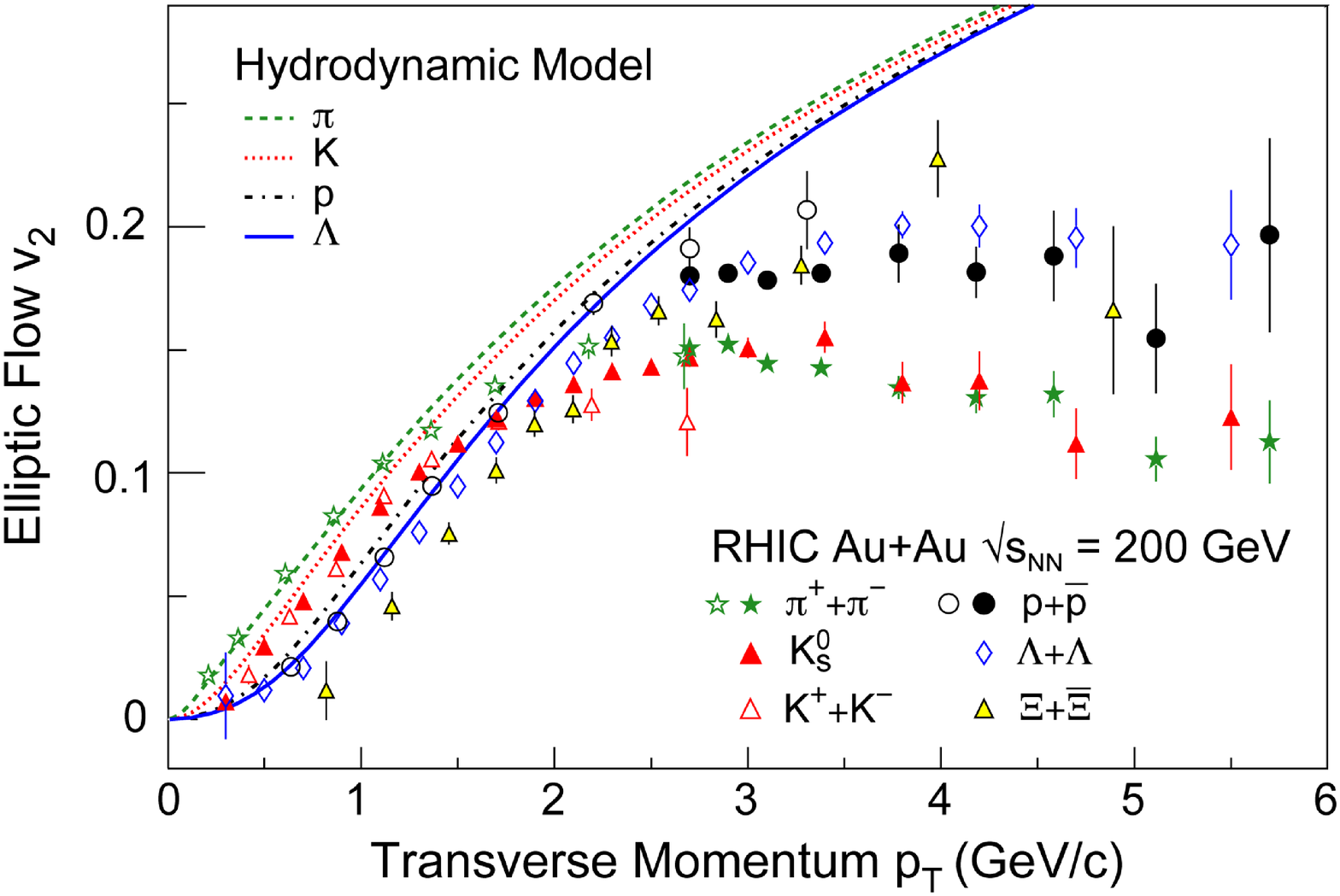}
\includegraphics[width=0.527\linewidth,clip=]{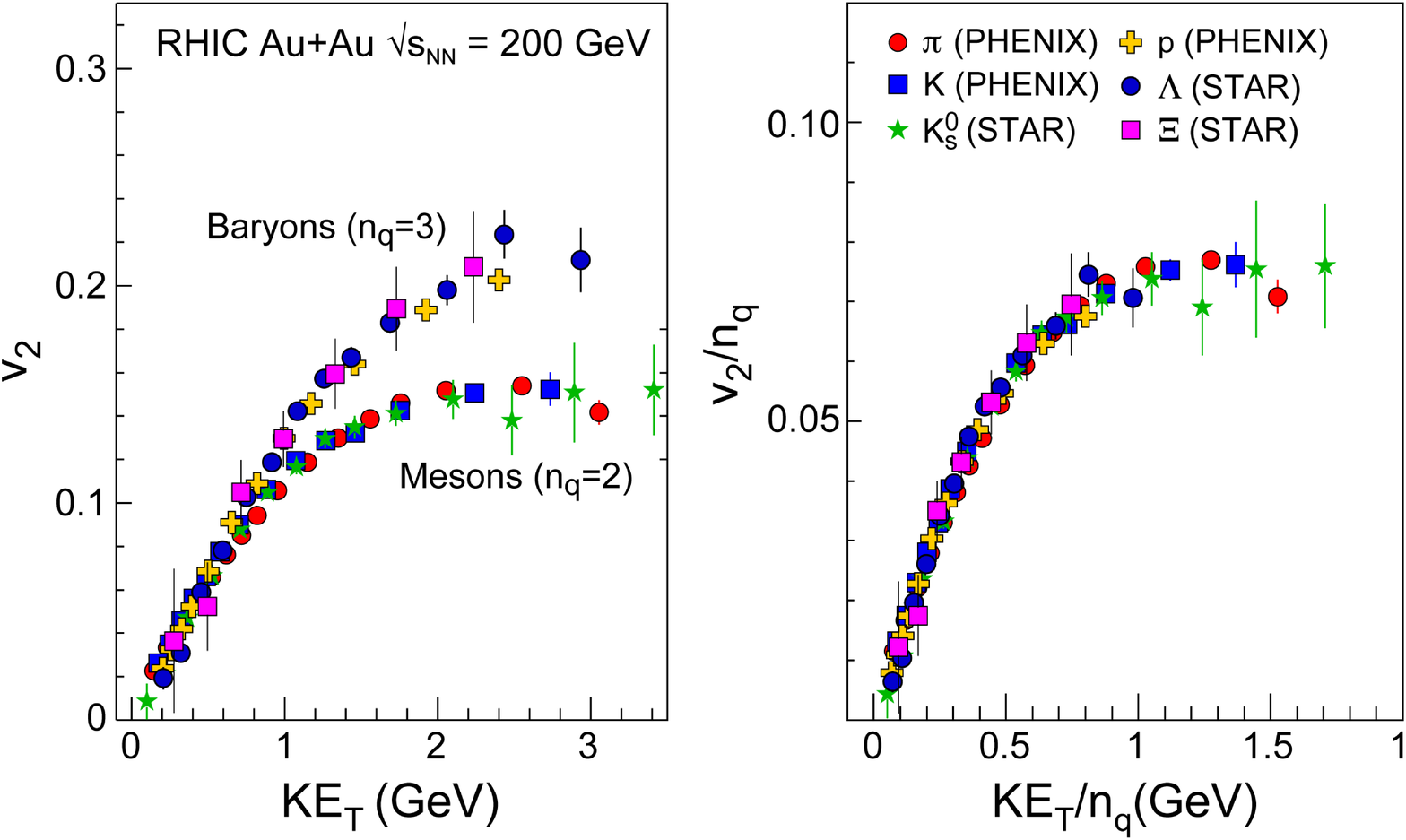}
\end{center}
\caption{\footnotesize
{\sl Left:} Up to $p_T\sim1.5$\,GeV/$c$, the differential elliptic flow
$v_2(p_T)$ follows the hydrodynamical predictions for an ideal fluid
perfectly \protect{\cite{STARv2}}. Note that $>99\%$ of all final hadrons
have $p_T<1.5$\,GeV/$c$.
{\sl Middle:} When plotted against transverse kinetic energy, the 
differential elliptic flow follows different universal curves for
mesons and baryons. 
{\sl Right:} When scaled by the number of valence quarks, the differential 
elliptic flow per quark follows the same universal curve {\em for all 
hadrons and for all values of (scaled) transverse kinetic energy} 
\protect{\cite{Adare:2006ti}}.
\label{F2}}
\end{figure}

Now let me return to the ``perfectly liquid QGP'' at RHIC. This is an
interesting story whose long version can be read in various review
articles (e.g. \cite{Muller:2006ee}). One chapter of this story is the
observation of ``jet quenching'', i.e. the suppression of high-$p_T$
jets by the dense medium created in the collision fireball. This is, of
course, something that excited Maurice tremendously, given his affi\-ni\-ty
to QCD jets throughout his life. I will here focus on a different part 
of the story, related to ``elliptic flow''. ``Elliptic flow'' denotes 
the anisotropic collective motion of particles produced in non-central 
heavy-ion collisions. If one pictures the two colliding nuclei as two 
spheres that smash into each other off-center, the part that actually 
interacts and gets stopped in the center-of-momentum frame is spatially
deformed like an egg or almond whose long axis is oriented perpendicular 
to the reaction plane. After thermalization of the fireball matter inside
that almond, the pressure gradient is larger in the short direction than 
in the long direction of the almond, so the matter is ejected with larger 
acceleration in the short direction. As a result, the final particles carry 
on average more momentum if they are parallel to the reaction plane than
perpendicular to it. One measures this by Fourier analysing the angular 
distribution around the beam direction of the finally observed hadrons
and extracting the second Fourier coefficient $v_2$. Figure~\ref{F2} 
shows this so-called ``elliptic flow coefficient'' for a variety of 
different hadron species, as a function of their transverse momentum $p_T$
or transverse kinetic energy $\mathrm{KE}_T=m_T-m_0=\sqrt{p_T^2+m_0^2}-m_0$.

Observation of elliptic flow is exciting because it unambiguously 
demonstrates final state interactions among the produced particles:
without interactions, there is no possibility to transform the initial
geometric deformation in coordinate space into a final momentum-space 
anisotropy. In fact, for a given spatial deformation, the largest 
momentum-space
anisotropy is expected for the most strongly interacting medium, i.e.
for the shortest mean free path of the produced particles. In the
limit of (approximately) zero mean free path, one expects thermalization
to occur most rapidly, resulting in almost perfect hydrodynamic collective
flow of the medium. Larger mean free paths would cause the fluid to be
more viscous and the resulting momentum anisotropy to be smaller
\cite{Romatschke:2007mq,Song:2007fn}. The largest possible final $v_2$
is thus predicted by ideal fluid dynamics. 

At the SPS, elliptic flow was measured in Pb+Pb collisions and found to be
about half of what ideal fluid dynamics predicts \cite{NA49v2}. But, as 
the left panel in Figure~\ref{F2} shows, at RHIC the experimental data 
exhaust the ideal fluid prediction! Deviations from the hydrodynamic 
prediction occur only for the small fraction ($<1\%$) of hadrons whose
transverse momenta are larger than $p_T > 1.5$\,GeV. Furthermore, the
data perfectly reproduce the hydrodynamically predicted dependence of 
$v_2(p_T)$ on the hadron rest masses. This thermal mass splitting can
be (approximately) absorbed by plotting $v_2$ as a function of transverse
kinetic energy instead of $p_T$ (middle panel in Fig.~\ref{F2}). Now
all curves collapse, not only in the low-$p_T$ region where hydrodynamics 
works, but even at higher transverse kinetic energies, where the 
hydrodynamic picture fails! To be precise, the data collapse onto two 
different curves, one for mesons and another for baryons. As if this were 
not miraculous enough, we can make these two branches overlap perfectly by
dividing both axes, the elliptic flow and the transverse kinetic energy, 
by the number of valence quarks in the hadron: now all hadrons fall
onto a single universal scaling curve (right panel)! What this tells
you is that the elliptic flow is carried by individual quarks, is the
same for the light up and down and the heavier strange quarks, and is 
transferred to the hadrons at the point of hadronization by quark
coalescence \cite{Greco:2003xt}. This is strong evidence for a key
dynamical role being played by deconfined quarks and antiquarks in
the QGP.

The observation of almost perfect fluid dynamical behaviour of the
bulk of the matter at low $p_T$ requires strong coupling in the QGP.
It is difficult and perhaps impossible to reproduce using QCD perturbation 
theory. But, for a special class of conformally symmetric quantum field 
theories (CFT), the strong-coupling limit can be analyzed with superstring 
theoretical methods, using the AdS/CFT duality \cite{Maldacena:1997re} which 
says that solving the strongly coupled CFT is equivalent to solving classical 
gravitational equations of motion in 5-dimensional curved space-times 
with anti-de-Sitter metric! In such theories, Son and collaborators found 
that the shear viscosity to entropy ratio becomes particularly small and
assumes a conjectured lower limit of $\frac{\eta}{s}=\frac{\hbar}{4\pi k_B}$
\cite{Policastro:2001yc}. A number of arguments, including some very
recent viscous hydrodynamic calculations of the elliptic flow
\cite{Romatschke:2007mq,Song:2007fn}, point strongly to an $\eta/s$ 
ratio for the QGP which is close to (i.e. not more than about a factor 
5 larger than) this conjectured lower limit. This makes the QGP the most 
perfect (real, not quantum) liquid ever observed in the laboratory! The 
only other system that seems to be able to get close is that of trapped 
ultracold atoms in the unitary (i.e. strongly interacting) regime 
\cite{coldatoms,Schafer:2007pr}.

As we moved from the BEVALAC via the SPS to RHIC, heavy-ion physics 
became mainstream. RHIC news now regularly captures the attention of the 
public and of scientists from other fields. Relativistic heavy-ion physics 
has always had a strong intellectual connection with astrophysics and 
cosmology (e.g. through quark stars, the conceptual similarities between 
the cosmological Big Bang and the Little Bangs created in heavy-ion 
collisions, and because RHIC explores the properties of the matter out 
of which our entire universe was initially made). As just noted, it  
now also has strong and fruitful bi-directional ties with the condensed 
matter physics of strongly coupled cold fermionic atoms, with high 
energy density physics through the physics of strongly coupled 
Coulomb plasmas, and with string theory and quantum gravity through the 
AdS/CFT correspondence which allows to explore strongly coupled 
gauge field theories with the help of their weakly coupled (classical) 
gravity duals. To foster such interdisciplinary ties has always been a 
hallmark of Maurice's work, at CERN and elsewhere. Naturally, he followed 
these developments with great enthusiasm until the end.     

{\bf Epilogue.} When I gave this talk I had just returned to 
CERN for a 1-year sabbatical. It was the first time for me at CERN 
without Maurice. I will always hold dear the spirited discussions I had 
with him during previous visits, and I will miss him this time around. But
his legacy remains with us. It takes visionaries like Maurice Jacob
to help create a young new field such as ultra-relativistic heavy-ion 
physics and nurture it to maturity. Those of us who have
heard him tell stories about his children know how proud and supportive 
Maurice has always been of them. He took the same care of his 
brain-children.

\smallskip 

{\bf Acknowledgements:} I thank the organizers of the ``Maurice Jacob Memorial
Meeting'', especially Luis Alvarez-Gaume, for giving me the opportunity to 
express my gratitude for what Maurice has done for my field of research, and 
Urs Wiedemann for helpful comments on the manuscript.


\end{document}